# In-Situ Neutron Diffraction and Crystal Plasticity Finite Element Modeling to study the Kinematic Stability of Retained Austenite in Bearing Steels

Rohit Voothaluru[a,*], Vikram Bedekar[a], Qingge Xie[b], Alexandru D. Stoica[b], R. Scott Hyde[a], Ke An[b]

[a]The Timken Company, North Canton, OH, 44720 USA

[b]Chemical and Engineering Materials Division, Spallation Neutron Source, Oak Ridge National Laboratory, Oak Ridge, TN 37831 USA

* Corresponding Author: Email Address: rohit.voothaluru@timken.com. Tel. +1-(234)-262-3984

## Abstract

This work integrates in-situ neutron diffraction and crystal plasticity finite element modeling to study the kinematic stability of retained austenite in high carbon bearing steels. The presence of a kinematically metastable retained austenite in bearing steels can significantly affect the macro-mechanical and micro-mechanical material response. Mechanical characterization of metastable austenite is a critical component in accurately capturing the micro-mechanical behavior under typical application loads. Traditional mechanical characterization techniques are unable to discretely quantify the micro-mechanical response of the austenite, and as a result, the computational predictions rely heavily on trial and error or qualitative descriptions of the austenite phase. In order to overcome this, in the present work, we use in-situ neutron diffraction of a uniaxial tension test of an A485 Grade 1 bearing steel specimen. The mechanical response determined from the neutron diffraction analysis was incorporated into a hybrid crystal plasticity finite element model that accounts for the martensite's crystal plasticity and the stress-assisted transformation from austenite to martensite in bearing steels. The modeling response was used to estimate the single crystal elastic constants of the austenite and martensite phases. The results show that using in-situ neutron diffraction, coupled with a crystal plasticity model, can successfully predict both the micro-mechanical and macro-mechanical responses of bearing steels while accounting for the martensitic transformation of the retained austenite.

**Keywords**: Transformation Induced Plasticity, Steel, Crystal Plasticity, In-situ neutron diffraction, Kinematic Stability

# 1. Introduction

High-carbon steels with microstructures composed of tempered martensite, retained austenite (RA) and carbides are prevalent in rolling element bearing applications. It is well known that the presence of retained austenite enhances ductility through the transformation-induced plasticity effect (TRIP). This phenomenon has been studied in detail by Voskamp [1], who examined the effect of load and loading cycles on the gradual decomposition of retained austenite and its subsequent effects on induced residual stress. Voskamp observed that the maximum transformation of retained austenite occurs subsurface, followed by maximum compressive stress at the corresponding depth. Voskamp et al. [2] later found that retained austenite transformation is a very sensitive parameter during contact fatigue and could play a major role in the elastic shakedown, steady state and instability stages of service life.

Since the TRIP effect enables retained austenite to increase ductility, several studies have been conducted on enhancing the retained austenite kinematic stability. Garcia-Mateo and Caballero [3] postulated that to maximize the benefits of retained austenite, its stability should neither be too high nor too low. They hypothesized that retained austenite with relatively low stability transforms too early in service life, causing the beneficial effects of TRIP transformation to remain unrealized. It was further stated that the presence of highly stable retained austenite at necking does not enhance ductility.

Bakshi et al. [4] developed a nanostructured bainitic steel by accelerating carbon migration into retained austenite and found that wear resistance during rolling/sliding conditions can be increased by increasing the retained austenite's stability. The study found that its stability depends not only on the amount of %C dissolved in the austenite, but also its size. Further, Xie et al. [5] conducted an in-situ EBSD study on austempered steel to understand the influence of size and shape on retained austenite stability. They found that film-like retained austenite transformed at a higher strain than the granular form. They created high-stability retained austenite by dissolving additional carbon and found that the retained austenite with higher stability gradually transformed through strain-induced martensitic transformation, which increased the work hardening index and delayed necking. Thus, it is well established that the mechanical stability of retained austenite is highly dependent upon chemistry (mainly its %C content), size and shape.

Blondé et al. [6] and Jimenez-Melero et al. [7] conducted very detailed thermo-mechanical studies of retained austenite stability using in-situ synchrotron radiations on TRIP and 52100 bearing steels. The authors reported the influence of temperature and load on retained austenite during uniaxial tensile testing. Their study on bearing steels was conducted on only one load (295MPa). A study on bearing steels using continuous loading has never been conducted, and since strain partitioning between the different constituents of steel also plays a role [8], such a study will enhance the body of knowledge that could be directly applied in industrial applications.

In order to quantify the rolling contact fatigue life of bearing components, it is imperative to understand the onset, steady state and complete transformation of retained austenite. While there is consensus regarding the beneficial effects of stable retained austenite in the service life of a bearing, the extent of this kinematic stability is yet to be qualified or quantified. A case in point is the fact that most of the tensile tests conducted in the prior research are limited to bulk samples with two or more dependent variables. Most of these studies assume that the strains in the retained austenite phase were identical to the macroscopically measured strains in the bulk. There is limited understanding of the actual stress and strain values at the onset of retained austenite transformation for bearing steels. Due to the lack of data with regards to the single crystal elastic constants of the austenite and martensite phases, most of the analytical and computational modeling of bearing steels is based upon continuum formulations using effective stress and strain data from the bulk macroscopic volumes of the bearing steel specimens.

The use of computational models to characterize these micro-mechanical responses is incomplete because there are very few accurate material models that can describe the difference in micro-mechanical response between the two individual phases present in bearing steel. The primary reason for the limited data availability in the literature is that retained austenite, being a metastable phase, cannot be studied in a discrete and independent manner without high-end instrumentation such as synchrotron radiation or neutron diffraction. However, the development of and recent advances in neutron sources have greatly enabled discrete studies at the lattice level that can help in understanding the transformation behavior of individual phases during tensile [9], torsion [10] and cyclic fatigue [11] testing. Neutrons penetrate deeper into the substrate, allowing characterization at the subsurface level that cannot be achieved using any other non-destructive techniques. The advantage of neutron diffraction at the Spallation Neutron Source facility is that the data can be acquired in real time without interrupting the tensile test, thus avoiding stress relaxation following the loading cycle. Also, the beam size of the neutrons is in the range of a few millimeters, compared to the micrometer range utilized by a synchrotron source. Using neutron diffraction, austenite transformation can be monitored in real time and in situ, while lattice parameters and the stability of the austenite can be studied simultaneously. Until now, most of the neutron diffraction studies have been performed on TRIP, stainless steel and non-ferrous materials, and little progress has been made on hardened and tempered (58-60 HRc) bearing steels. In this study, a discrete and in-situ analysis of retained austenite transformation is conducted using high-carbon (1% wt) bearing steel and by employing a state-of-the-art engineering neutron diffractometer [10]. The study sheds light on deformation dynamics and transformation behavior based on the lattice strains experienced by the austenite and martensitic planes during continuous loading.

The ubiquity of bearings in industrial applications puts a greater emphasis on studying and understanding the potential micro-mechanical response in material microstructures. To achieve this, there is a need to develop computational models that can accurately model the micro-mechanical response of the individual steel phases. This will help in quantifying the elastic constants of the individual phases viz. retained austenite and martensite in bearing steels. Over the past decade, crystal plasticity finite element (CPFE) models have been successful in predicting micro-mechanical response and also in estimating the fatigue life or relative fatigue life of various polycrystalline microstructures. Recent advances in high-performance computing have helped significantly to enhance the use of CPFE models to predict both the short- and long-term effects of application loads in different representative volumes. Manonukul and Dunne [12] were among the first to use CPFE models to predict low cycle fatigue in nickel alloys. Several groups have developed fatigue initiation parameters (FIP) using CPFE models to facilitate a relative comparison between different application loads [13-15]. Recently, Voothaluru and Liu [16] used CPFE models to predict the micro-mechanical response and also the potential fatigue life of copper [17] and iron microstructures [18]. Alley and Neu [19, 20] developed a hybrid crystal plasticity formulation to predict rolling contact fatigue life in high-carbon steels. Alley and Neu found that CPFE models can capture the macro-mechanical response of bearing steels very well. They also found that hybrid crystal plasticity models can accurately capture the effect of retained austenite on the macro-scale material response. Recently, Woo et al. [21, 22] used CPFE models to predict the micro-mechanical response of individual ferrite and martensite phases in a dual phase steel. In order to successfully understand and model the micro-mechanical response of bearing steels, there is need to quantify the single crystal elastic constants of the material using CPFE models as they can successfully capture the lattice strain response with good agreement. So far, there has been little progress in the development of quantified material models for CPFE modeling of high-carbon bearing steels that account for transformation-induced plasticity. This is primarily due to the difficulty of mechanically characterizing the kinematically metastable retained austenite. As a result, the majority of the computational models handling bearings in the industry are still heavily reliant on continuum mechanics formulations. Continuum mechanics models and damage mechanics models, while computationally inexpensive, are not capable of accurately capturing the effect of the presence of heterogeneity in microstructures in bearing steels with multiple phase constituents.

In order to address this issue, we present an in-situ neutron diffraction based on an empirically quantified material model for CPFE modeling of multi-phase high-carbon bearing steel. The results of the lattice strain response from in-situ neutron diffraction were used to develop a material model for the bearing steel using CPFE modeling based on a hybrid constitutive formulation that was developed based upon the works of Asaro [23], Turteltaub and Suiker [24], and Alley and Neu [19]. The computational framework and modeling schema follow the works of Voothaluru and Liu [16]. The CPFE model was implemented using a user material subroutine (UMAT) in ABAQUS.

## 2. In-Situ Neutron Diffraction – Experimental Details
### 2.1 Sample Characterization

The sample used in the current study is a dogbone specimen of A485 Grade 1 (A485-1) steel, the composition of which is listed in Table 1. AISI A485-1 with a slightly higher Si content was chosen instead of standard through-hardened 52100 steel, as Si is known to stabilize the retained austenite. The samples were austenitized at $850^{o}C$ for 45 minutes, followed by quenching in water. Subsequent tempering was conducted at $180^{o}C$ for 1.3 hours. The hardness was measured to be 62.4 HRc. Microstructural characterization was carried out using scanning electron microscopy and was found to be composed of tempered martensite, retained austenite and carbides. The retained austenite content was found to be 18%. The %C was found to be 0.9% in the retained austenite, calculated at Proto Inc. using the technique described by Lason et al. [25].

| C  | Mn    | Si   | Cr    | Ni    | P      | S      | Fe  |
|----|-------|------|-------|-------|--------|--------|-----|
| 1% | 1.09% | 0.6% | 1.06% | 0.11% | 0.013% | 0.012% | Bal |

**Table 1**: Chemical Composition of A485-1 Steel (wt.%)

## 2.2 Neutron Diffraction

An in-situ neutron diffraction experiment was conducted to determine the lattice strains under uniaxial loading, as shown in Fig. 1. The significantly large penetration depth of the neutrons and the volume-averaged nature of the bulk measurement that is characteristic of a scattering beam are very well suited for understanding deformation behavior in polycrystalline materials [26]. Dogbone-shaped tensile test samples with a cylindrical cross section were heat-treated and machined for the neutron diffraction and tension testing. The gauge length of the samples was 115mm; their diameter was 6.35mm. The in-situ neutron diffraction experiments were conducted on the VULCAN engineering diffractometer [10] at Spallation Neutron Source (SNS) in the Oak Ridge National Laboratory (ORNL). The VULCAN time of flight (TOF) diffractometer enabled rapid collection of structural changes in the sample under dynamic loading conditions. The schematic for the in-situ setup during mechanical loading has been discussed elsewhere [27].

A 45° angle between the sample and the incident neutron beam was maintained. Two detector banks (Bank 1 and Bank 2) located at ±90° from the incident beam collected data from the longitudinal (LD) and transverse (TD) directions, respectively. The neutron beam size was 5mm x 5mm and the collimator size was 5mm, enabling data collection over a 125mm$^3$ gauge volume. Prior to loading, a reference scan was collected for 10 minutes (longer than the bin time) to minimize the $d_0^{hkl}$ propagated statistical error [28]. The sample was continuously loaded at room temperature until the elastic load of 28kN was reached. The loading continued until fracture. This continuous loading eliminated stress relaxation under the stress/strain control during holding [10]. The neutron diffraction data was collected simultaneously and in real time during the loading. The data was analyzed using two-minute interval bins with the event-based software VDRIVE (Vulcan Data Reduction and Interactive Visualization Software) [29, 30] and the peak position in d spacing was fitted by performing a single peak fit. The lattice strains were calculated using eq. (1), where $\varepsilon^{hkl}$ is the lattice strain, $d_0^{hkl}$ and $d^{hkl}$ are the before and after strains.

$$\varepsilon^{hkl} = \frac{d^{hkl}-d_0^{hkl}}{d_0^{hkl}} \tag{1}$$

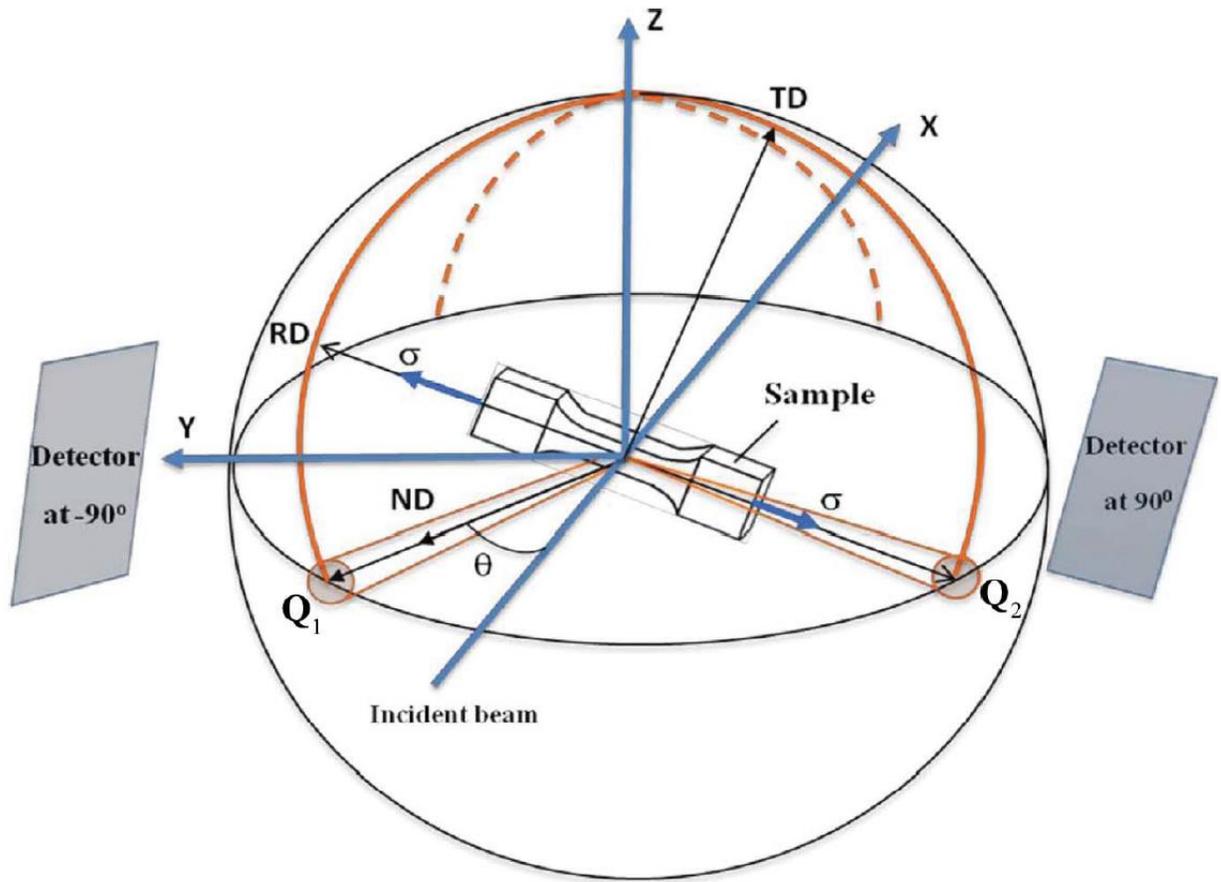

Fig. 1: Schematic for the in situ measurements (loading) at the VULCAN diffractometer: XYZ is the instrument coordinate system, xyz is the sample coordinate system (identified, respectively, with the main processing directions, rolling, RD, transverse, TD, and normal, ND), $\sigma$ is the mechanical loading along the RD, $\theta$ is the diffraction angle, and Q1 and Q2 are the scattering vectors [27]

## 3. Crystal Plasticity Finite Element Modeling

Computational modeling of high-carbon steels using continuum mechanics and linear elastic fracture mechanics assumptions limits the use of the models primarily to simplistic macroscopic analyses. In order to understand the mechanical response of multiphase steel accurately, it is imperative to model the plastic behavior that is controlled by microplasticity. In the case of high-carbon steels, the model must also account retained austenite and martensite crystal plasticity and also the stress-assisted transformation from austenite to martensite. This will help in quantifying the single crystal elastic constants of the individual phases present in A485-1 Steel. Crystal plasticity models have been developed successfully for simulating the behavior of bearing steels over the past decade. However, most modeling in this field has been driven predominantly by phenomenological formulations or use micromechanics approaches to qualify the macro-mechanical responses. In the present work, we are particularly interested in developing a material modeling approach that relies on a coupled empirical input and modeling prediction that allows us to accurately quantify the micro-mechanical response of the representative volume element (RVE). The mathematical formulation for the crystal plasticity model demonstrated here is based upon the work of Asaro [23], which focuses on a rate-dependent model with a multiplicative decomposition of the deformation gradient. The transformation from austenite to martensite is accounted in the model following the stress-assisted transformation models developed and implemented by Turteltaub and Suiker [24]. A detailed description of the constitutive model can be found in previous papers by Turteltaub and Suiker [24] and Alley and Neu [19]. The polycrystal plasticity framework is implemented using a User Material Subroutine (UMAT) in ABAQUS following the works of Alley and Neu [20] and Voothaluru and Liu [18].

The total deformation gradient $\boldsymbol{F}$ is given by:

$$\boldsymbol{F} = \boldsymbol{F}^e . \boldsymbol{F}^p . \boldsymbol{F}^{tr} \qquad (2)$$

where, $F^{tr}$ is the transformation gradient that accounts for the volumetric strain produced by austenite-martensite phase transformation, $F^p$ accounts for the polycrystalline plasticity in the martensite and $F^e$ is the elastic deformation gradient.

The model is built using this deformation gradient to simulate the plasticity and phase transformation that occur along slip and transformation systems associated with the lattice structures of the martensite and austenite. The model assumes that the 48 BCC slip systems will exhibit behavior approximately similar to that of the BCT tempered martensite. The critical stress for transformation and microplasticity are based on the Cauchy stress tensor, $\sigma$, so the model assumes that transformation is given priority since its critical threshold is lower than the crystal plasticity of martensite [19]. The UMAT follows a two-step procedure for modeling the combined behavior of transformation and plasticity in ABAQUS. The total deformation gradient at the beginning and end of each time step was input, and the tangent modulus was determined.

The transformation strain increment within each step was determined first via an iterative Newton-Raphson method. The plastic deformation gradient did not vary during this step. Subsequently, the stress was recalculated and the plastic deformation iterated to balance the external load. The two-phase formulation was incorporated into an ABAQUS UMAT, which follows an implicit integration algorithm. The model evaluates the shear states and transformation rates at the end of the given time step. The plastic shearing rate $\dot{\gamma}^{(\alpha)}$ on the $\alpha^{th}$ slip system is governed by the rate-dependent flow rule:

$$\dot{\gamma}^{(\alpha)} = \dot{\gamma}_0 \cdot \left| \frac{\tau^{(\alpha)} - \chi^\alpha}{g^{(\alpha)}} \right|^m \cdot sgn(\tau^{(\alpha)} - \chi^{(\alpha)}) \qquad (3)$$

where, $m$ is the strain rate sensitivity exponent, $\dot{\gamma}_0$ is the shearing rate coefficient, $g^{(\alpha)}$ is the drag stress, $\chi^{(\alpha)}$ is the back stress and $\tau^{(\alpha)}$ is the resolved shear stress on the $\alpha^{th}$ slip system. The resolved shear stress on each slip system is related to the Cauchy stress tensor, according to:

$$\tau^\alpha = \sigma : (s^{(\alpha)} \otimes m^{(\alpha)}) \qquad (4)$$

The drag stress and back stress evolution follows the expressions in Eq. (5) and (6), respectively:

$$\dot{g}^{(\alpha)} = \sum_{\beta=1}^{N_{slip}} H_{dir} \cdot |\dot{\gamma}^{(\beta)}| - g^{(\alpha)} \cdot \sum_{\beta=1}^{N_{slip}} H_{dyn} \cdot |\dot{\gamma}^{(\beta)}| \quad (5)$$

$$\dot{\chi}^{(\alpha)} = A_{dir} \cdot \dot{\gamma}^{(\alpha)} - \chi^{(\alpha)} \cdot A_{dyn} \cdot |\dot{\gamma}^{(\alpha)}| \quad (6)$$

where, $H_{dir}$ is the isotropic hardening coefficient, $H_{dyn}$ is the dynamic recovery coefficient and $A_{dyn}$ is the dynamic recovery coefficient for the back stress.

The rate of volume fraction transformation $\dot{\xi}^{(\lambda)}$ on a transformation system $\lambda$ is given by:

$$\dot{\xi}^{(\lambda)} = \dot{\xi}_{max} \cdot \tanh\left(\frac{1}{v^{tr}} \cdot \left(\frac{<f_{tr}^{\lambda} - f_{cr}^{\lambda}>}{f_{cr}^{\lambda}}\right)\right) \quad (7)$$

where, $\dot{\xi}_{max}$ is the maximum rate of transformation, $v^{tr}$ the viscosity parameter, $f_{cr}^{\lambda}$ the critical driving stress and $f_{tr}^{\lambda}$ is the driving stress on the $\lambda^{th}$ transformation system. The driving stress on $\lambda^{th}$ transformation system is related to the transformation and habit vectors $\hat{b}^{\lambda}$ and $\hat{n}^{\lambda}$ and the Cauchy stress tensor by:

$$f_{tr}^{\lambda} = \boldsymbol{\sigma}:(\gamma_T \cdot \hat{b}^{\lambda} \otimes \hat{n}^{\lambda}) \quad (8)$$

where $\gamma_T$ is the shape strain magnitude, a parameter that is uniform for all transformation systems. The rate of change of the volume fraction transformed from austenite to martensite is given by Eq. (9), where $\dot{V}_{trans}$ is the rate of transformation of retained austenite.

$$\dot{V}_{trans} = \sum_{\lambda=1}^{N_{trans}} \dot{\xi}^{(\lambda)} \quad (9)$$

The critical driving force $f_{cr}^{\lambda}$ is controlled by the transform rates, as shown in Eq. (10). Here, Q is the transform hardening coefficient along transformation plane $\eta$. This accounts for the increased resistance to transformation as more of the retained austenite becomes surrounded by transformed martensite. The increase in resistance to transform on any system is assumed to be the same in line of previous works by Alley and Neu [19].

$$f_{cr}^\lambda = \sum_{\eta=1}^{N_{trans}} Q \cdot |\dot{\xi}^{(\eta)}| \qquad (10)$$

The transformation gradient that accounts for the volumetric transformation from austenite to martensite is given by:

$$\boldsymbol{F}^{tr} = \sum_{\lambda=1}^{N} \gamma_T \xi^{(\alpha)} \widehat{\boldsymbol{b}}^\alpha \otimes \boldsymbol{m}^{(\alpha)} \qquad (11)$$

The transformation is active only when the Macaulay brackets are satisfied in Eq. (7). The transformation is also unidirectional on each system and $\xi^{(\lambda)}$ is non-negative. There are 24 transformation systems for the austenite, of which 12 are reverse vectors of the others. This configuration was employed to ensure that although transformation can occur in either direction, the austenite would only transform into martensite and not the reverse [19].

Prior works on high carbon bearing steels did not account for the transformation of the FCC retained austenite and the deformation of the product martensite discretely. In order to account for this, in the present work, the FCC retained austenite was modeled using the same constitutive formulation however, the material constants were calibrated following the development of the martensitic model constants. The FCC retained austenite has 12 slip systems so the resulting shearing rate was calculated using the constitutive model over the 12 FCC slip systems. In addition, the constitutive model was completed by specifying the evolution of its elastic modulus with martensitic volume fraction, which is expressed by eq. (12) where, $\boldsymbol{C}_{RA}$ and $\boldsymbol{C}_M$ are the moduli of austenite and the product martensite, whose crystallographic orientation is aligned with that of the parent austenite phase.

$$\boldsymbol{C} = (1 - \xi_M)\boldsymbol{C}_{RA} + \xi_M \boldsymbol{C}_M \qquad (12)$$

The simulations were run in ABAQUS 6.14 using RVEs with a grain size parameter that generated a randomized distribution microstructure model, and with a fixed average grain size of $10\,\mu$m for the aggregate as per the experimental data.

## 4. Results
### 4.1 Experimental Results
#### 4.1a Scanning Electron Microscopy and X-ray Diffraction

SEM analysis was conducted using a Versa 3D FIB/SEM microscope. Fig. 2 shows the microstructure composed of tempered martensite, retained austenite and carbides. X-ray diffraction analysis was conducted using Proto LXRD equipment with Chromium ka radiation of wavelength 2.28Å. The retained austenite content was found to be 18%.

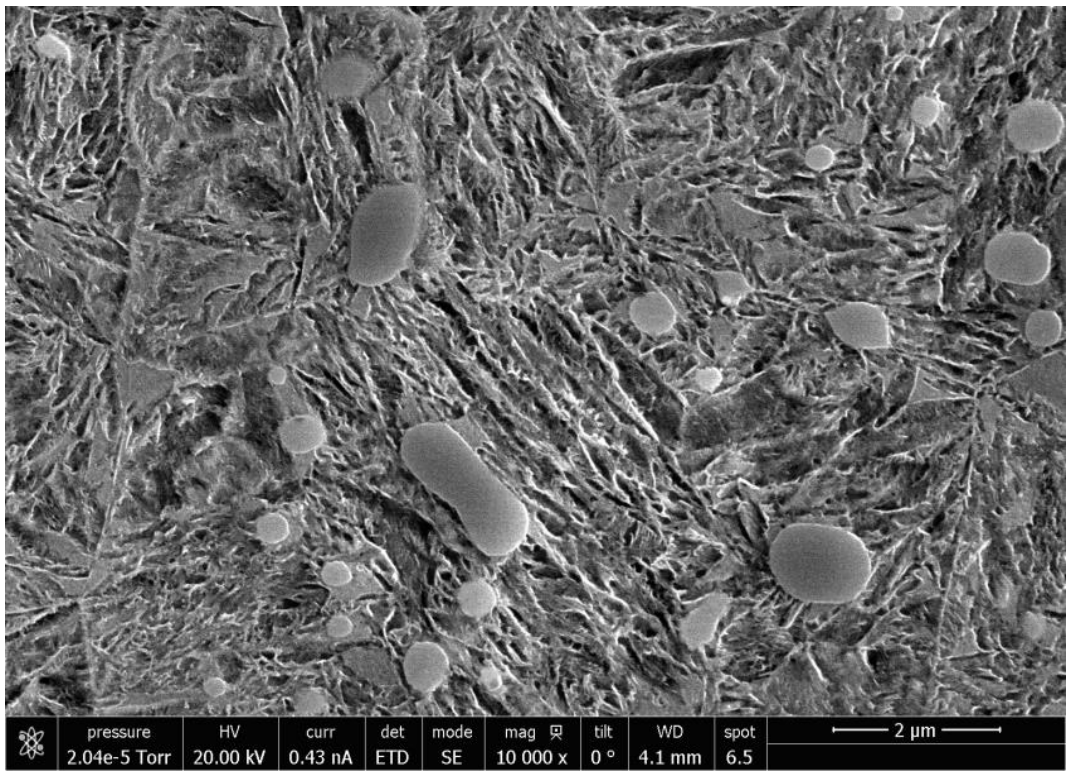

Fig. 2: SEM image showing microstructure of heat-treated A485 Grade 1 steel

#### 4.1b Neutron Diffraction

The bulk macro-mechanical response of the A485-1 steel is shown in Fig. 3(a). The sample indicated yield strength of 1.057GPa at true strain of 0.006. The elastic-plastic portion of the deformation was also recorded and a fracture stress of 1.9GPa was observed for this sample. The 3D representation of the intensity, with respect to the loading time, for austenite {200} planes (d = 1.8Å) is seen in Fig. 3(b), while the 2D representation of the same is seen in Fig. 3(c). The intensity plot (Figs. 3(b) and 3(c)) shows that the retained austenite was relatively stable when the deformation was within the elastic limit. Subsequently, the transformation was observed around the yield point (as indicated by arrows). The austenite transformation continued until fracture stress was reached.

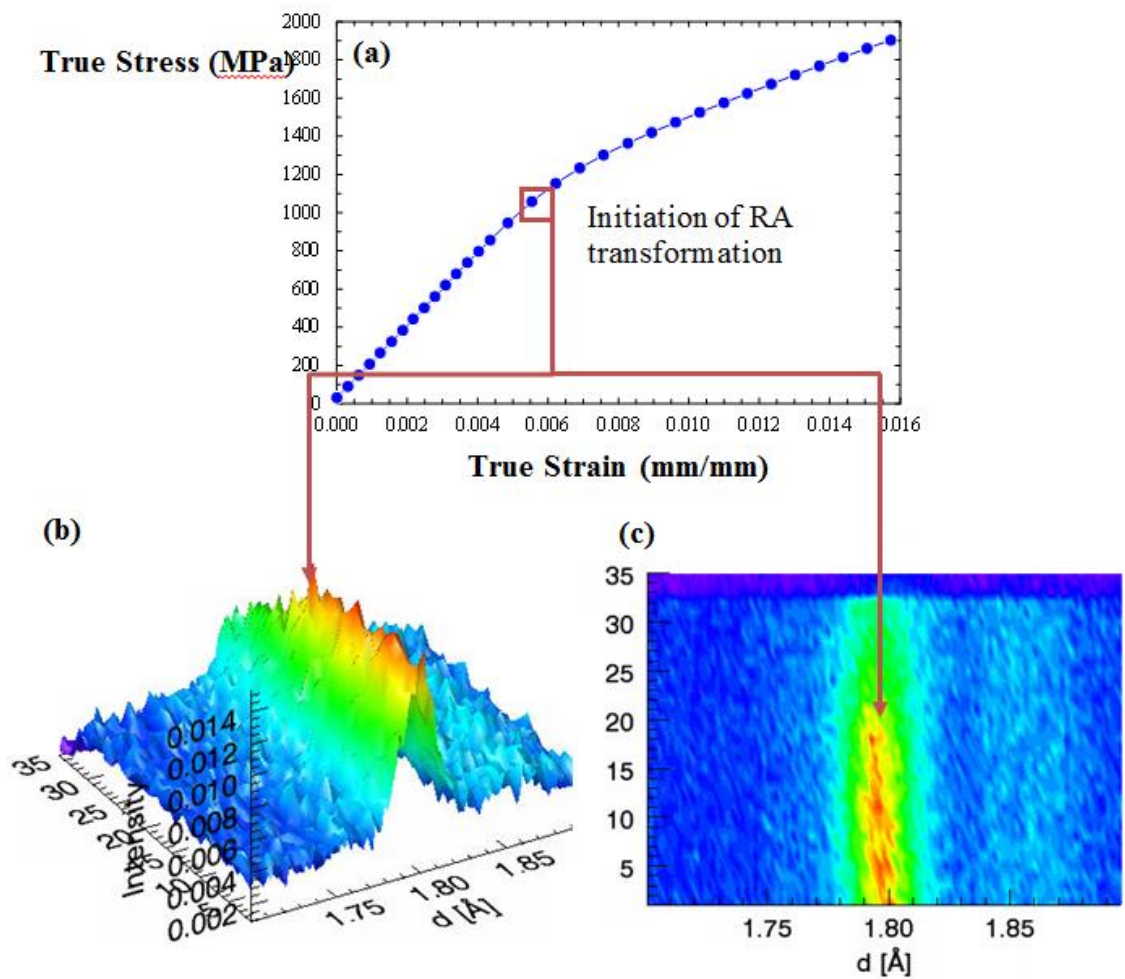

Fig. 3: (a) Macroscopic (true) stress-strain behavior of A485-1 steel (b) 3D time-dependent intensity for austenite {200} planes; (c) 2D time-dependent intensity for austenite {200} planes

Fig. 4 further details the lattice stress-strain curve for three martensite (BCC) and austenite (FCC) planes. It is evident that the retained austenite remained stable until 1.057GPa, which was also the macroscopic yield strength. As the stress reached a critical value, the mechanical stimuli necessary for TRIP transformation were met and the transformation was suddenly triggered on all austenite planes. The critical value of stress is perhaps dependent upon a complex interplay between the grain size, %C in austenite and the surrounding matrix. Interestingly, the transformation of the retained austenite also played a key role in determining the yield point of the bulk sample. Beyond the yield stress, the strain in the austenitic planes flattens, suggesting that the austenite planes began to slip and could not take further stress. It should be noted that the standard deviation of the lattice strain was observed to be relatively low. Specifically for martensite {110} planes, it ranged from 0.03% to 0.07%.

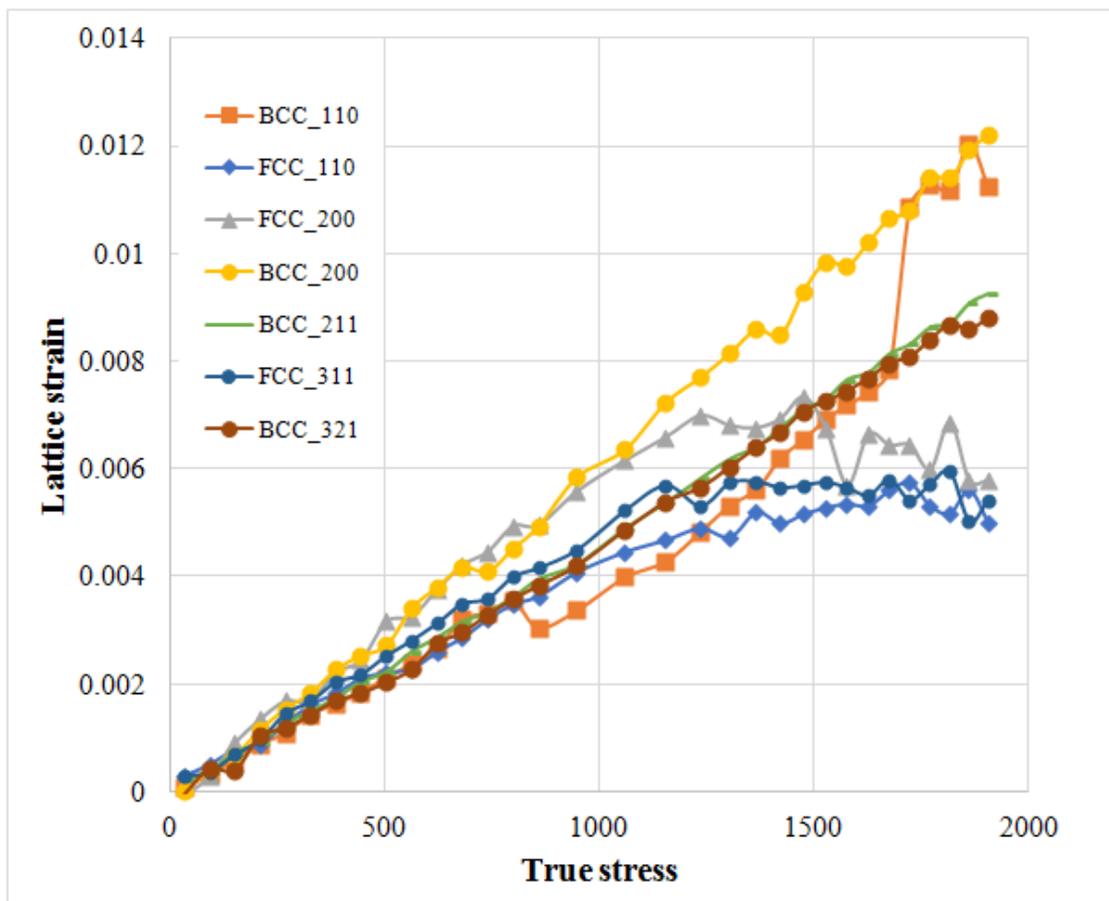

Fig. 4: Lattice strain estimation from in-situ neutron diffraction of uniaxial tension testing

## 4.2 Modeling Results
### 4.2.1 Determination of Microscopic Material Parameters

The polycrystal plasticity model follows a constitutive formulation that is reliant on a semi-empirical data validation scheme. The model needs empirically driven inputs to validate the independent variables and the rate lattice strain in the austenite and martensite phases. In order to achieve this, the polycrystal plasticity model was first calibrated by comparing the micro-mechanical response of the RVE with the corresponding empirical data for the macroscopic and lattice stress-strain curves. The computational model which, is an RVE with C3D8R elements, was developed using an RVE generator. Periodic boundary conditions were applied to the RVE along the surfaces parallel to the loading axis. Fig. 5(a) and 5(b) illustrate the RVE Model in ABAQUS along with the boundary conditions used for the current analysis and the austenite and martensite portions of the model.

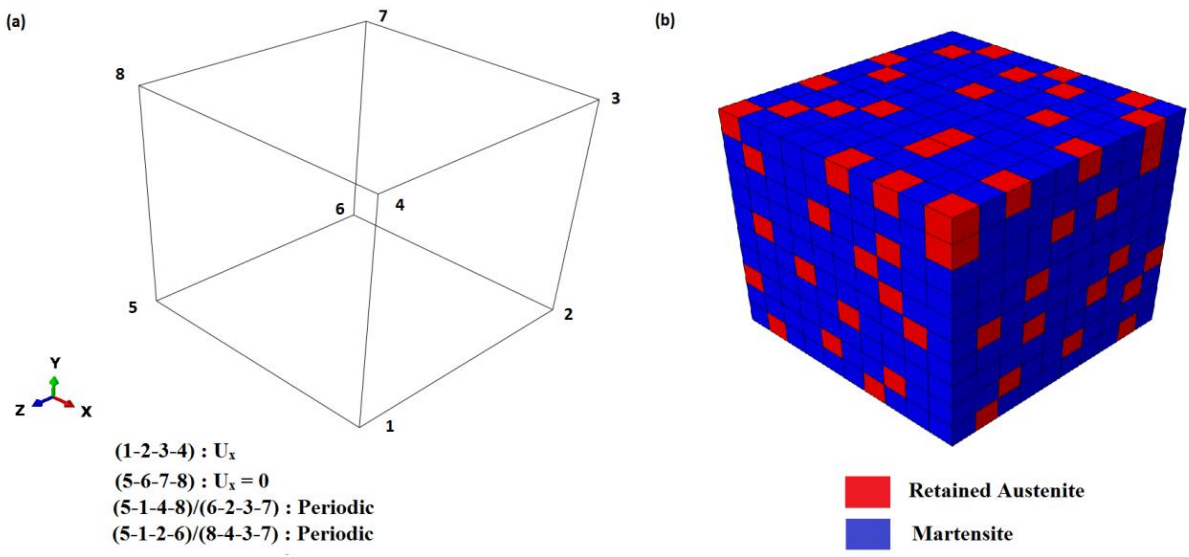

Fig. 5 (a): RVE with boundary conditions (b) showing retained austenite and martensite phases

In order to allow the non-uniform response of the different orientations, a simple prescribed displacement boundary condition of the faces would lead to artificial responses along those surfaces. In order to allow for localized deformations and measure the overall response of the grains collectively, periodic boundary conditions are applied to the model as described in Smit et al. [31], Alley and Neu [20] and Voothaluru and Liu [18]. The surface nodes on face 1-2-3-4 and 5-6-7-8 were tied to two different reference points and the reference node was subject to displacement boundary conditions. The reference node tied to face 5-6-7-8 was fixed and the one that was tied with 1-2-3-4 face was applied the displacement incrementally. Periodic boundary conditions are applied to the lateral faces. The model was set up using 1000 elements (10x10x10) with an initial volume fraction of 18% retained austenite. The grains with austenite were identified using a RVE generator that tags the austenite grains in the matrix. 50 RVE aggregates were generated in this fashion. The location of the austenite grains and martensite grains were randomly tagged as a part of the RVE generator. The predicted result of the lattice strain and macroscopic stress-strain curve is an average of the mechanical response predicted across the 50 RVEs generated in this manner. The formulation of the two-phase model relies on constants being fitted for both the crystal plasticity and transformation plasticity parts of the model. The calibration process was carried out in two steps. First, the A485-1 steel response that was obtained from uniaxial deformation experiments was used to calibrate the single-phase crystal plasticity model. While this calibration was carried out, the transformation model was bypassed and the apparent mechanical response in this stage was assumed to represent the martensite phase behavior. For the purely martensitic samples, the lattice strain response was not studied and only the macro-mechanical loading and stress-strain data were used to calibrate the material model. The elastic constants and the material parameters used to get the appropriate fit are listed in Table 2. The initial shearing rate coefficient was set to $0.001s^{-1}$, which is typical for martensitic crystal plasticity. The strain rate sensitivity exponent (*m*) was set to 50 to simulate near rate-independent behavior. For this model, the material was assumed to be demonstrating isotropic hardening, and as a result, the initial back stress and the kinematic hardening coefficient were set to zero. The dynamic recovery coefficients for drag and back stress were also set to zero. Fig. 6 illustrates the effective strain computed for one RVE model after uniaxial loading boundary conditions were simulated. The calibrated model data for the purely martensitic response assumption is shown in Fig. 7. The material parameters obtained in this fashion are assumed to represent the behavior of the martensite phase in the two-phase material model as well. The transformation plasticity and the volumetric change incorporated into the austenite-martensite transformation model were calibrated using the lattice stress-strain data collected using neutron diffraction. The values of the shape strain magnitude ($\gamma_T$), transform viscosity parameter ($v_{tr}$) and maximum rate of transformation ($\dot{\xi}_{max}$) in the austenite transform model were set to 0.1809, 0.17 and 0.003 s$^{-1}$, respectively [24].

Table 2: Material Model Parameters (GPa)

| Martensite | | | | | Austenite | | | | |
|---|---|---|---|---|---|---|---|---|---|
| C11 | C12 | C44 | $H_{dir}$ | $g_0^{(\alpha)}$ | C11 | C12 | C14 | $H_{dir}$ | $g_0^{(\alpha)}$ |
| 278.7 | 114.2 | 90.2 | 6.9 | 0.84 | 229.1 | 101.2 | 85.4 | 6.4 | 0.56 |

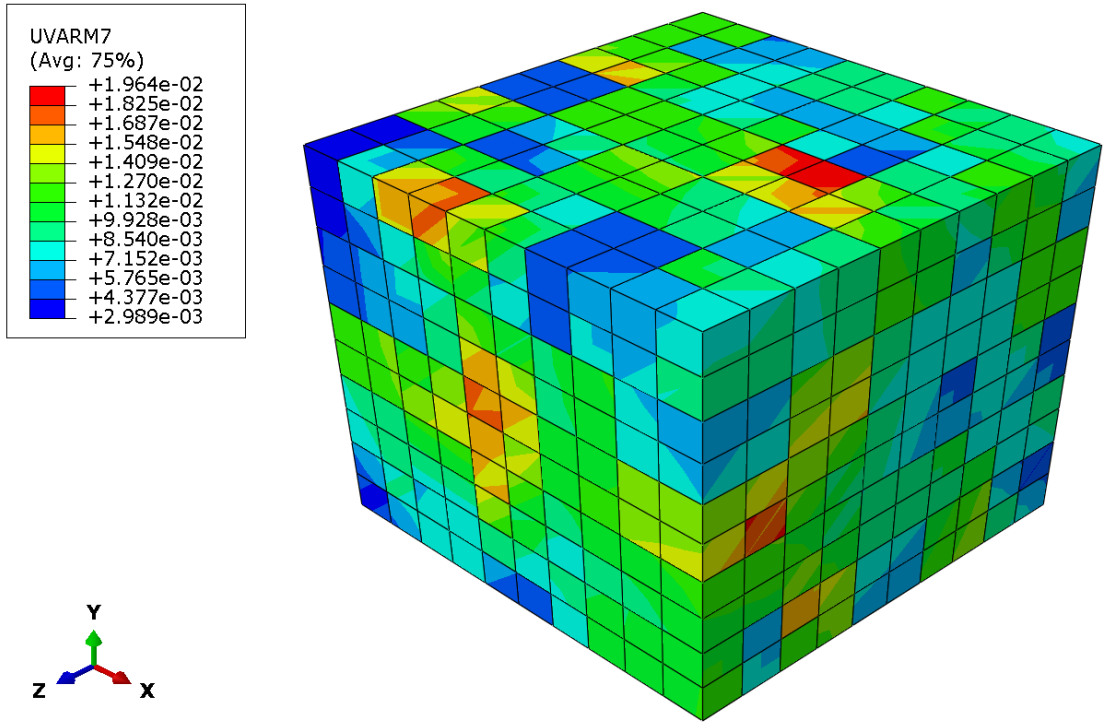

Fig. 6: RVE model in ABAQUS and the effective strain after uniaxial loading

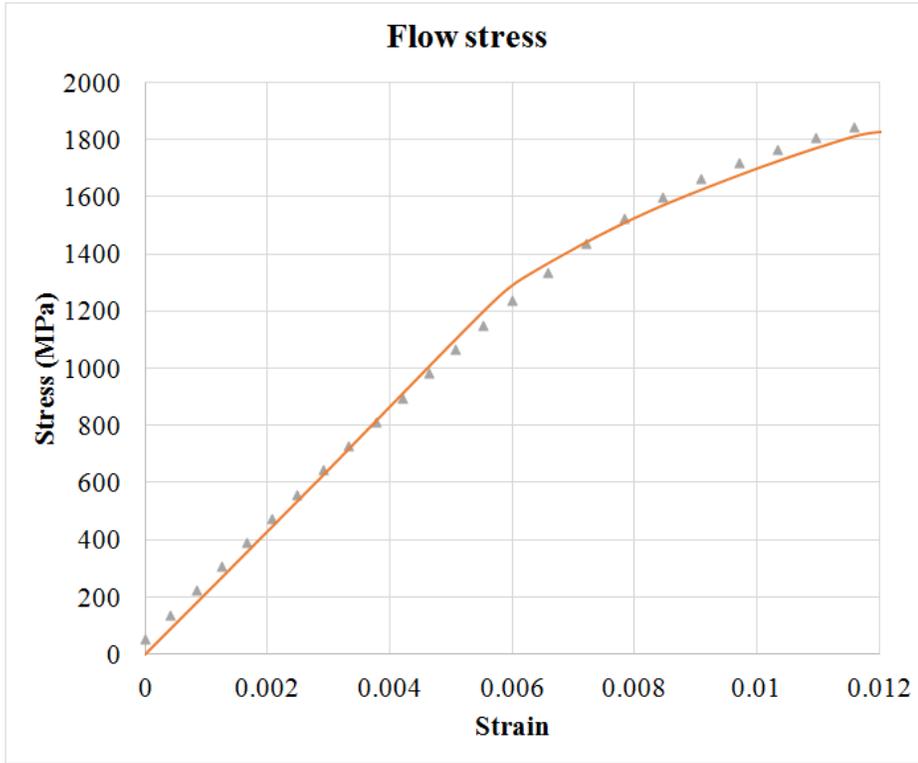

Fig. 7: Stress-strain response of fully martensitic bearing steel – model parameter fit

**4.2.2 Micro-mechanical Behavior Modeling**

Using the material parameters obtained from the microscopic parameter fit, in Table 2 and 3, the model was simulated with the boundary conditions discussed above. The micro-mechanical hardening parameters were fit iteratively to match the lattice stress-strain response and the macro-mechanical response. Two parameters — critical driving stress, $f_{cr}^{\lambda}$, and direct hardening parameter of transformation, Q — were varied following generation of the initial trial values as discussed in detail in prior works by Alley and Neu [19]. The resulting material parameters that allow the best fit are listed in Table 2 and Table 3. The results for the lattice strain along <hkl> directions parallel to the loading direction were evaluated from the computational model. From the model prediction for lattice strains shown in Figs. 8 and 9, we can observe that once the austenite starts transforming, the volumetric strain due to the transformation results in a slight increase in the lattice strain of the martensite phase as well due to the response from the product martensite. This prediction is in line with the empirical data, as we can see that beyond the yield point, the martensitic portion of the sample shows a slightly increasing gradient in the rate at which the lattice strain is accumulating along the (200) direction. This matches the experimental results reasonably well. The minor deviations from the empirical data are believed to be a combination of the effect of the accuracy of the data collected and the averaged computation from the RVE aggregates.

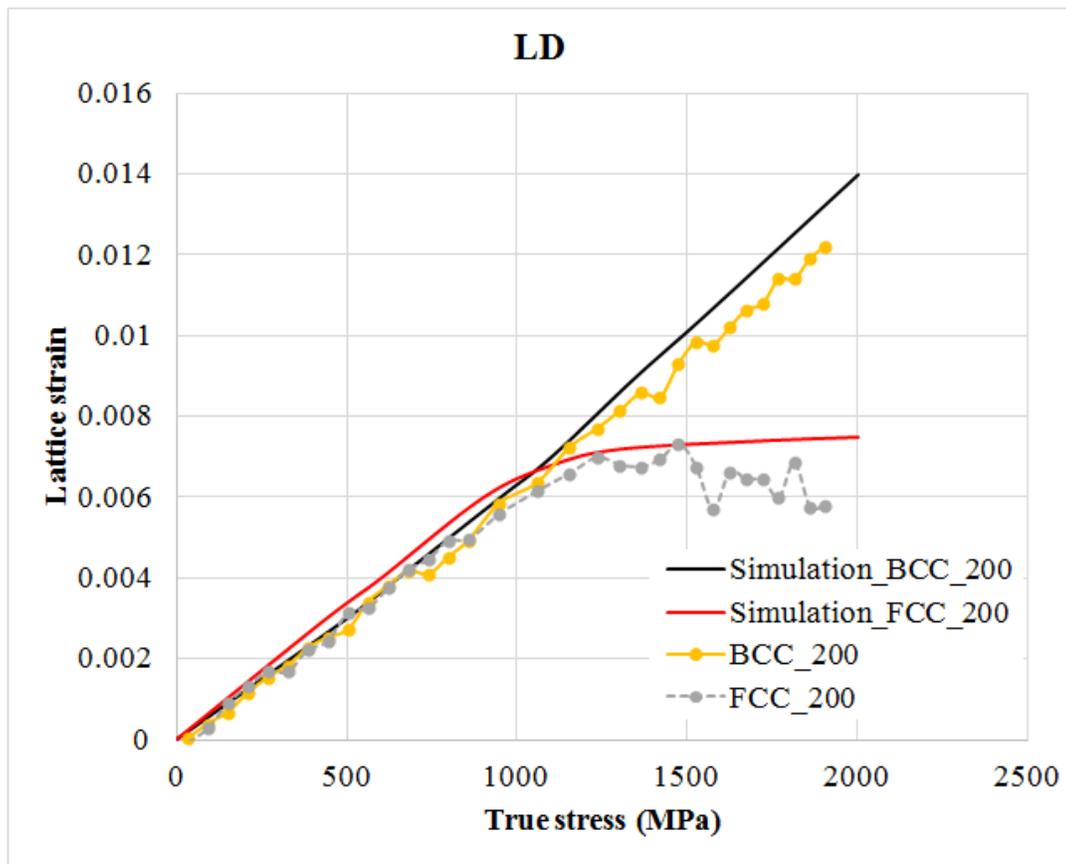

Fig. 8: Comparison of the lattice strain parallel to the (200) direction predicted by the model with the empirical data from in-situ neutron diffraction during uniaxial tension testing

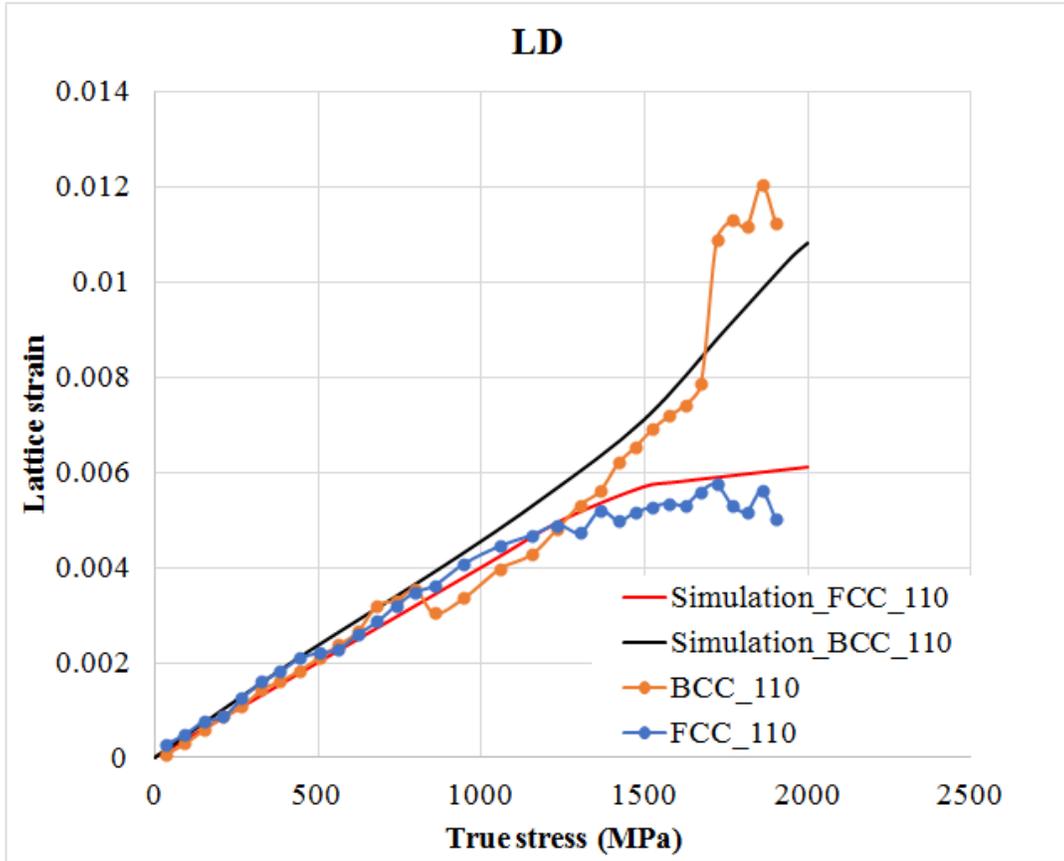

Fig. 9: Comparison of the lattice strain parallel to the (110) direction predicted by the model with the empirical data from in-situ neutron diffraction during uniaxial tension testing

Table 3: Transformation Model Parameters

| $RA_{initial}$ | $\gamma_T$ | $v_{tr}$ | $\dot{\xi}_{max}$ | $f_{cr}^{\lambda}$ | $Q$ |
|---|---|---|---|---|---|
| 18% | 0.1809 | 0.17 | 0.003s$^{-1}$ | 97.1MPa | 549MPa |

## 5. Discussion

The kinematic stability of the retained austenite in A485-1 bearing steel was evaluated systematically using in-situ neutron diffraction of an uniaxial tension test. The effect of the retained austenite stability on the macroscopic stress strain response and lattice strain evolution at room temperature was studied and a crystal plasticity model was used to quantify the elastic and micro-plastic hardening parameters for this steel. The transformation behavior of retained austenite observed in this study was unique when compared with the reported works in existing literature. Until now, transformation behavior using neutron diffraction has mainly been studied using low-carbon TRIP steel. In those cases, the retained austenite transformed progressively well ahead of the macroscopic yield stress [6]. Besides grain size and %C, one of the major differences between the TRIP steel and the bearing steels is the surrounding microstructure. In TRIP steel, the matrix microstructure is composed of softer allotriomorphic ferrite, while in the case of bearings steels, the surrounding microstructure is composed of very hard tempered martensite and carbides. Thus, it is believed that the matrix microstructure might be playing an important role in the stability and transformation of the retained austenite. As reported in a recent work by Bedekar et al. [32], preferential transformation along austenite {200} planes was observed with a consequent increase in the martensite {211} planes. This could be due to the crystal symmetry rules, as reported by Drahokoupil et al. [33]. The decrease in the intensity for {200} was more dramatic compared to the rest of the planes studied. This could be due to the highest number of active slip planes exhibited by {200} planes accommodating the lattice strain. Overall, the lattice stress-strain data presented in this study indicate a unique phenomenon related to retained austenite transformation. The highest strain was accommodated by martensite. The austenite undertook less stress as the load was transferred to the martensite. The data also indicate complexities of retained austenite transformation with preferred transformation along the {200} planes.

In order to computationally capture the micromechanical response of the material in this work and to estimate the elastic and micro-plastic hardening constants, a crystal plasticity finite element model based computational study was carried out. The model used a two step material parameter calibration approach that allowed us to capture the onset of phase transformation as observed from the change in the micro-mechanical response of the austenitic phase from linearity. The semi-empirical nature of this constitutive formulation allowed for modifying the critical driving stress $f_{cr}^{\lambda}$ and the direct hardening parameter of transformation $Q$ in an iterative fashion to fit the micro-mechanical response of the RVE aggregate. The results also show that the modeling formulation with the parameter fit is capable of capturing the lattice strain response in both the austenite and martensite phases reasonably. The austenite transformation started around a lattice strain of 0.006 in grains parallel to the (200) direction, and it can be observed that the volumetric strains due to the transformation are causing the deviation from linearity for the micro-mechanical response of the martensite phase. The resulting material parameters were then used to predict the macroscopic mechanical response of the A485-1 steel RVE. Fig. 10 shows the stress-strain curve predicted using the updated material parameters from Table 3. The results have shown that using the hybrid crystal plasticity formulation, coupled with two empirical inputs (viz. one for martensitic steel and one for a two-phase steel with quantified amount of retained austenite), we can accurately capture the micro-mechanical and macro-scale response of bearing steels.

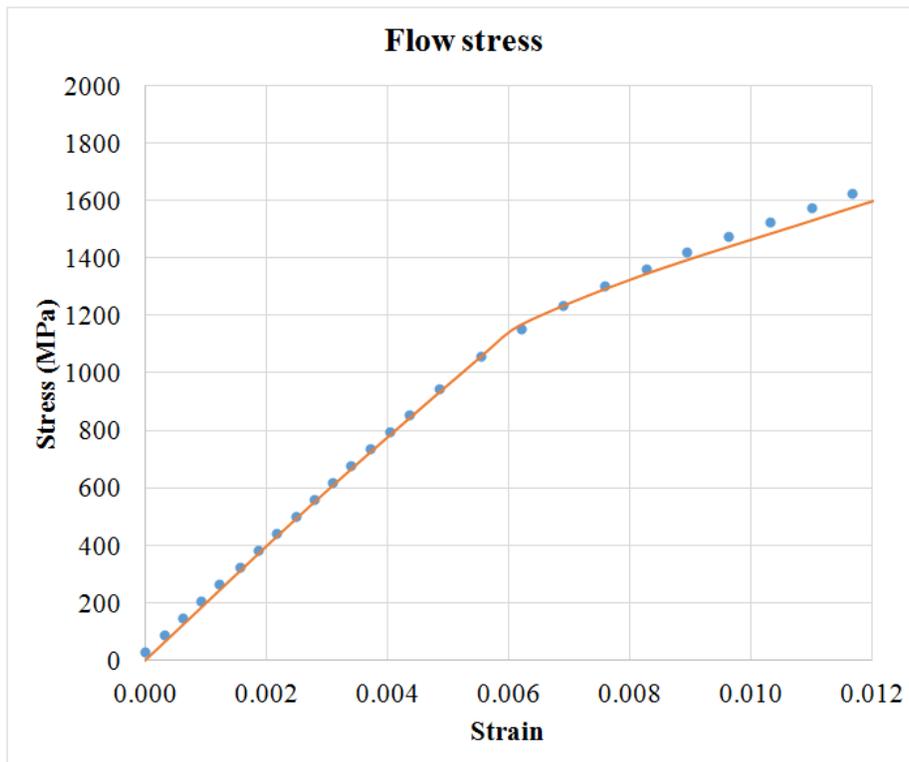

Fig. 10: Comparison of the predicted macroscopic stress-strain response from the model with the empirical estimation observed from uniaxial tension testing

From the micro-mechanical response of the bearing steel in this model and experiment, we can see that the kinematic stability of the retained austenite phase can be reasonably estimated by coupling the neutron diffraction data with a CPFE formulation. The CPFE model allowed us to quantify and determine the microscopic hardening parameters for the bearing steel under consideration. From the results we could observe that the austenite within bearing steels is relatively stable until it reaches a threshold strain which is corresponding to a true stress about 1.1GPa. This is very much in line with the experimental results which show the deviation from linearity starts at around the yield point.

## 6. Conclusion

The present work demonstrated the use of a CPFE model in predicting the micro-mechanical and macro-mechanical responses of dual phase high-carbon bearing steels. The material model for the CPFE model was developed and validated using empirical data for the lattice strain response of individual phases in the bearing steel, using in-situ neutron diffraction of uniaxial tension testing of through-hardened steel specimens. The results have shown that CPFE models coupled with an empirical technique such as in-situ neutron diffraction can result in very good predictive capability for quantifying the microstructural response of bearing steels. The microscopic hardening parameters and single crystal elastic constants of the martensite and retained austenite phases in A485-1 bearing steel were determined by modeling the elastic and elastic-plastic portions separately using a two-step parameter fit approach for martensite and austenite phases. The resulting predictions from the model matched the empirically observed trends very well.

In addition, the neutron diffraction experiment has shown that the retained austenite in AISI A485-1 bearing steel was uniquely stable in the macro-scale elastic regime. It was also observed that the retained austenite TRIP transformation was triggered almost at the same time, since there was detectable plastic strain from the macro-scale test. The kinematic stability of retained austenite within bearing steels is of prime importance to bearing applications and this work enabled us to understand the kinematic stability of the retained austenite while simultaneously estimating the material parameters that would allow future studies to be conducted computationally.

**Acknowledgments**

The authors thank Dr. Stephen P. Johnson (Director, Timken Technology Center) for permission and support for this work. We gratefully acknowledge the support and guidance of Dr. Richard W. Neu (Professor, Georgia Institute of Technology) during the development of the UMAT. We would also like to thank Mr. Matt Boyle, Mr. Chris Akey, Mr. Robert Pendergrass and associates in the Prototype department for sample preparation.

A portion (neutron diffraction) of this research used resources at the Spallation Neutron Source, a DOE Office of Science User Facility operated by the Oak Ridge National Laboratory (ORNL).